\documentclass[aip,jcp,reprint,
floatfix,
]{revtex4-1}
\usepackage{xcolor}
\usepackage{amsmath}
\usepackage{graphicx}
\usepackage{dcolumn}
\usepackage{bm}

\usepackage[utf8]{inputenc}
\usepackage[T1]{fontenc}
\usepackage{mathptmx}

\begin{document}


\title{Effect of dissolved salt on the anomalies of water at negative pressure}

\author{Alberto Zaragoza}\thanks{These authors contributed equally.}
\affiliation{Departamento Estructura de la materia, f\'{i}sica t\'{e}rmica y electr\'{o}nica , Facultad de Ciencias F\'{i}sicas, Universidad Complutense de Madrid, 28040 Madrid, Spain}
\affiliation{Departamento de Ingenier\'{i}a F\'{i}sica, Divisi\'{o}n de Ciencias e Ingenier\'{i}as, Universidad de Guanajuato,37150 Le\'{o}n, M\'{e}xico}
\author{Chandra Shekhar Pati Tripathi}\thanks{These authors contributed equally.}
\affiliation{Universit\'e de Lyon, Universit\'e Claude Bernard Lyon 1, CNRS, Institut Lumi\`ere Mati\`ere, F-69622, Villeurbanne, France}\thanks{CSPT now at: Department  of  Physics,  Institute  of  Science,  Banaras  Hindu  University,  Varanasi-221005, India.}
\author{Miguel A.Gonzalez}
\author{Jos\'{e} Luis F. Abascal}
\affiliation{Departamento de Qu\'{i}mica F\'{i}sica I, Facultad de Ciencias Qu\'{i}micas, Universidad Complutense de Madrid, 28040 Madrid, Spain}
\author{Fr\'{e}d\'{e}ric Caupin}\email{frederic.caupin@univ-lyon1.fr}
\affiliation{Universit\'e de Lyon, Universit\'e Claude Bernard Lyon 1, CNRS, Institut Lumi\`ere Mati\`ere, F-69622, Villeurbanne, France}
\author{Chantal Valeriani}
\affiliation{Departamento Estructura de la materia, f\'{i}sica t\'{e}rmica y electr\'{o}nica , Facultad de Ciencias F\'{i}sicas, Universidad Complutense de Madrid, 28040 Madrid, Spain}

\date{20 April 2020}

\begin{abstract}
Adding salt to water at ambient pressure affects its thermodynamic properties. At low salt concentration, anomalies such as the density maximum are shifted to lower temperature, while at large enough salt concentration they cannot be observed any more. Here we investigate the effect of salt on an anomaly recently observed in pure water at negative pressure: the existence of a sound velocity minimum along isochores. We compare experiments and simulations for an aqueous solution of sodium chloride with molality around $1.2\,\mathrm{mol\,kg^{-1}}$, reaching pressures beyond $-100\,\mathrm{MPa}$. We also discuss the origin of the minima in the sound velocity and emphasize the importance of the relative position of the temperatures of sound velocity and density anomalies.
\end{abstract}

\maketitle

\section{Introduction}

While being the most familiar liquid, water is also the most peculiar. It exhibits thermodynamic anomalies, such as a density maximum near $4\,^{\circ}\mathrm{C}$ and a compressibility minimum near $46\,^{\circ}\mathrm{C}$ at ambient pressure. It also exhibits dynamic anomalies at low temperature\cite{singh_pressure_2017}, with the shear viscosity decreasing and the self-diffusion coefficient increasing with applied pressure. Among several theoretical scenarios proposed to explain the origin of water's anomalies (see Ref.~\onlinecite{gallo_water_2016} for a review), the second critical point scenario\cite{poole_phase_1992} postulates the existence of a phase transition between two distinct metastable liquids terminating at a liquid-liquid critical point (LLCP). One key feature of the second critical point scenario is the existence of a maximum in isothermal compressibility along isobars, observed in several molecular dynamics simulations of water\cite{xu_relation_2005,pi_anomalies_2009,abascal_widom_2010,abascal_note_2011,gonzalez_comprehensive_2016,biddle_two-structure_2017,kim_maxima_2017}. We note that such a maximum could also exist without a liquid-liquid transition: in the singularity free conjecture~\cite{sastry_singularity-free_1996}, a maximum may arise as a thermodynamic consequence of the existence of density anomalies; an example is provided in models where the LLCP is at zero temperature\cite{stokely_effect_2010,Anisimov_thermodynamics_2018}.
Obtaining experimental evidence for a compressibility maximum is a challenging task because it is predicted to lie at large supercooling, very close or even beyond the homogeneous ice nucleation line~\cite{holten_equation_2014}, which is an experimental limit for supercooling real water.

To bypass this limitation, a program was started at the University of Lyon to measure the equation of state of supercooled water at negative pressure. Negative pressure is another metastable state of water, with respect to vapor. Metastability can be terminated by the nucleation of a bubble. However, it has long been recognized~\cite{roedder_metastable_1967,Zheng_liquids_1991} that micron-sized fluid inclusions (FI) in quartz provide samples of sufficient cleanliness to reach very large negative pressures, beyond -100 MPa, and close to the homogeneous cavitation limit~\cite{caupin_liquid-vapor_2005,el_mekki_azouzi_coherent_2013,Menzl_molecular_2016} (see also Refs.~\onlinecite{caupin_cavitation_2006,Caupin_escaping_2015} for reviews on cavitation in water).

The transparent FI samples can be probed by light scattering techniques. In particular, Brillouin spectroscopy, based on the inelastic interaction between light and density fluctuations in a material, can be used to measure sound velocity $c$. Following this approach, Pallares \textit{et al.} first observed the existence of minima in $c$ vs. temperature along the path followed by a FI during cooling\cite{pallares_anomalies_2014}. For a perfectly rigid quartz matrix, the FI would follow an isochore; due to thermal expansion and elasticity of quartz, the liquid density varies slightly and the actual path is referred to as a quasi-isochore. Based on the observed $c$ minima and on a comparison with molecular dynamics simulations, it was proposed that the $c$ minima along quasi-isochores were related to isothermal compressibility $\kappa_T$ maxima along isobars. This long-sought anomaly would become accessible at negative pressure because it would emerge at temperatures above that of homogeneous ice nucleation (which has since been measured at negative pressure using FIs in quartz~\cite{qiu_exploration_2016}). Further work with other FIs at several densities~\cite{holten_compressibility_2017} confirmed the existence of sound velocity minima. Interpolation of the sound velocity data allowed, by thermodynamic integration, to reconstruct the equation of state of water at negative pressure~\cite{pallares_equation_2016,holten_compressibility_2017}. This showed that the temperature of density maxima, increases less and less rapidly with negative pressure, reaching $18.2\,^{\circ}\mathrm{C}$ at -137 MPa~\cite{holten_compressibility_2017}. Remarkably, the most recent data~\cite{holten_compressibility_2017} suggests the existence of a line of $\kappa_T$ maxima along isobars at the edge of the experimentally accessible region (around $-10\,^{\circ}\mathrm{C}$ and $-100\,\mathrm{MPa}$). This work also delineated the causal relations between the different lines of anomalies and their relative order. In particular, thermodynamics require that the line of $c$ minima must lie closer to the line of density maxima than the the line of $\kappa_T$ maxima. This means that the line of $c$ minima is easier to access experimentally before ice nucleation occurs.

Two months after publication of Ref.~\onlinecite{holten_compressibility_2017}, Kim \textit{et al.} published another report of a maximum in $\kappa_T$, this time near zero pressure~\cite{kim_maxima_2017}. They used fast evaporation of small droplets in vacuum to prepare liquid water at temperatures lower than usual, and pulses from an x-ray laser to measure the static structure factor $S(q)$ at wavenumber $q$. Extrapolating $S(q)$ to $q=0$ yields a quantity proportional to $\kappa_T$. Kim \textit{et al.} concluded that a $\kappa_T$ maximum exists at $229\,\mathrm{K}$. However, extracting $\kappa_T$ requires calculating the droplet temperature, and an extrapolation of density data. This extrapolation of density data and the existence of a $\kappa_T$ maximum have been later debated~\cite{caupin_comment_2018,Kim_Response_2018}, and the accuracy of the reported temperature put into question~\cite{goy_shrinking_2018,caupin_thermodynamics_2019}. Nevertheless, if it exists, the $\kappa_T$ maximum near zero pressure would be compatible with the $\kappa_T$ maxima reported at negative pressure. We also note that a recent 2-state model for water \cite{caupin_thermodynamics_2019} which assumes the existence of a LLCP is able to quantitatively reproduce many thermodynamic measurements both at positive and negative pressure.

In order to shed light on the origin of water's anomalies, more data are highly needed. One possible route is to add a solute to water. This has the effect of modifying the phase diagram and moving or suppressing the lines of anomalies. A solute also shifts the ice nucleation line to lower temperatures. A good candidate is salt (sodium chloride NaCl). Its effect on the anomalies of water has been studied in experiments~\cite{archer_thermodynamic_2000} and simulations~\cite{corradini_route_2010,corradini_liquidliquid_2011}. It is predicted that the LLCP seen in simulations of pure water shifts to higher temperature and lower, possibly negative, pressure. As the line of $\kappa_T$ maxima emanates from the LLCP, it is also shifted in the same direction. However, the line of density maxima is shifted to lower temperatures. Based on thermodynamic reasoning~\cite{holten_compressibility_2017}, a line of $c$ minima must exist between the lines of $\kappa_T$ and density maxima in the solution. However, its location with respect to the line of $c$ minima in pure water cannot be predicted by thermodynamic arguments only. Here we address this question by a combined experimental and simulation study of an aqueous solution of sodium chloride at negative pressure. Numerical and experimental results are presented in Sections~\ref{sec:sim} and \ref{sec:exp}, respectively. They are then compared in Section~\ref{sec:discuss}, where we discuss the possible origins of sound velocity minima.
	
\section{Numerical results\label{sec:sim}}

\subsection{Simulation details}
Molecular dynamics simulations were carried out with GROMACS 2016.4~\cite{hess_gromacs_2008}, using 91 Na$^{+}$ and Cl$^{-}$ ions and 3818 water molecules, which corresponds to a molality of $1.323\,\mathrm{mol\,kg^{-1}}$. Water was simulated using the TIP4P/2005 model~\cite{abascal_general_2005} which consists of one Lennard-Jones and three Coulombic sites, while NaCl ions were simulated using two different models: Joung-Cheatham \cite{joung_determination_2008} and the so called Madrid model \cite{benavides_potential_2017}. Water-ions interactions are reported in table \ref{forcefields}.
When simulating water, we truncated the Lennard-Jones (LJ) potential at 9.5~\AA, adding standard long-range corrections to the LJ energy, and
using Ewald sums (with PME technique)\cite{essmann_smooth_1995} for the calculation of the long-range electrostatic forces, with a real space cut-off at 9.5~\AA. Periodic boundary conditions were applied in all directions.  We set the time step to $1\,\mathrm{fs}$ and simulate every temperature for at least $200\,\mathrm{ns}$ (lower temperatures required longer simulations reaching up to $500\,\mathrm{ns}$). In order to ensure proper equilibration, we checked at every state point that no drift was detectable in any thermodynamic property, such as the energy. In order to keep temperature and pressure constant, we used a Nose-Hoover thermostat~\cite{nose_unified_1984} and a Parrinello-Rahman barostat ~\cite{parrinello_polymorphic_1981} with a relaxation time set to $1\,\mathrm{ps}$ for both.

\begin{table}[ttt]
	\caption{Interaction parameters for the force-fields used in this work: Joung-Cheatham - TIP4P/2005, and Madrid Model - TIP4P/2005}.
	\label{forcefields}
  \begin{tabular*}{0.48\textwidth}{@{\extracolsep{\fill}}ccc}

\hline  
\hline  
        & Joung-Cheatham - TIP4P/2005 &  \\
\hline
        & $\sigma/\mathrm{nm}$ & $\epsilon/(\mathrm{kJ\,mol^{-1}})$\\
\hline

       Na$^{+}$~-Na$^{+}$ & 0.2159538        & 1.475465   \\
       Cl$^{-}$-Cl$^{-}$ & 0.4830453        & 0.053493   \\
       Na$^{+}$-Cl$^{-}$ & 0.3494996       &  0.2809396  \\
       O-O               & 0.315890         & 0.774907   \\
       Na$^{+}$-O        & 0.2659219        & 1.069275   \\
       Cl$^{-}$-O        & 0.399445         & 0.2035979  \\

\hline
\hline
	& Madrid model - TIP4P/2005 &  \\
\hline
        & $\sigma/\mathrm{nm}$ & $\epsilon/(\mathrm{kJ\,mol^{-1}})$\\
         \hline
	Na$^{+}$-Na$^{+}$ & 0.221737         & 1.472360     \\ 
	Cl$^{-}$-Cl$^{-}$ & 0.484906         & 0.076923   \\
	Na$^{+}$-Cl$^{-}$ & 0.290512         & 1.438890   \\
	O-O               & 0.315890         & 0.774907   \\
	Na$^{+}$-O        & 0.251338         & 0.793388   \\
	Cl$^{-}$-O        & 0.426867         & 0.061983   \\

	\hline 
          & Madrid model - Charges/(e) &	\\
    \hline
	q$_{Na^{+}}$ = -q$_{Cl^{+}}$   & &  0.85    \\ 
	q$_{H}$ = -q$_{M}$/2          &    &  0.5564   \\
\hline
\hline 
\end{tabular*}
\end{table} 

We have performed $NVT$ (where $N$ is the number of particles, $V$ the volume, and $T$ the temperature) simulations to calculate pressure $P$ and isochoric heat capacity $C_{V}$, using:
	\begin{equation}\label{ec_heat_capacity_volume}
		C_{V}=\frac{\left< U^{2}\right> -\left< U\right>^{2}}{k_{B}T^{2}},
	\end{equation}
where $U$ is the energy and $k_{B}$ is the Boltzman constant. We have also performed $NPT$ simulations to calculate isobaric heat capacity $C_{P}$ and isothermal compressibility $\kappa_{T}$, using:
	\begin{eqnarray}
	C_{P} & = &\frac{\left< H^{2}\right> -\left< H\right>^{2}}{k_{B}T^{2}}, \label{ec_heat_capacity_pressure}\\
	\kappa_{T} & = &\frac{\left< V^{2}\right> -\left< V\right>^{2}}{\left< V\right> k_{B}T},\label{ec_isothermal_compressibilty}
	\end{eqnarray}
where $H$ is the enthalpy. Finally, the speed of sound $c$ was obtained via the  Newton-Laplace formula:
	\begin{equation}\label{ec_speed_of_sound1}
	c = \sqrt{\frac{C_{p}/C_{V}}{\kappa_T \rho}},
	\end{equation}
	where $\rho$ is the mass density of the liquid. To simulate a system at a given density $\rho$, the appropriate box size is set in $NVT$ simulations, and, in $NPT$ simulations, the barostat is set at the pressure measured in $NVT$ simulations at the desired $\rho$. We have computed error bars for $C_V$, $C_p$, and $\kappa_T$ via the block average method with 8 blocks. Errors on sound velocity (obtained through error propagation) and on pressure were found to be smaller than the figures’ symbols.

\subsection{Comparison between Joung-Cheatham and Madrid models}\label{sec:compar_models}

\begin{figure}[bbb]
\begin{center}
\includegraphics[clip,width=0.95\columnwidth]{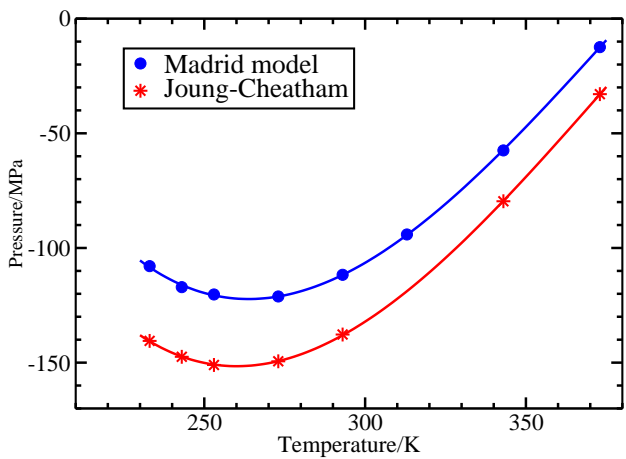}
	\end{center}
	\caption{Pressure as a function of temperature along the isochore at $\rho=996.5\,\mathrm{kg\,m^{-3}}$ for Joung-Cheatham (red stars) and Madrid (blue circles) models at $1.323\,\mathrm{mol\,kg^{-1}}$. Error bars are smaller than the symbol size.}
\label{modelspvst}
\end{figure}

We first compare the pressure computed along the $\rho=996.5\,\mathrm{kg\,m^{-3}}$ isochore 
for the two models (Fig.~\ref{modelspvst}). This density value was chosen to obtain a pressure with the Madrid model close to the one estimated for the experiment (see Section~\ref{sec:exp}). The two models give similar results, with the Madrid model yielding a pressure shifted up by around $30\,\mathrm{MPa}$ compared to the Joung-Cheatham model. Both isochores show a minimum around $260-263\,\mathrm{K}$, which corresponds to a point along the line of density maxima of the solution. The data for each extrema observed in the simulations are given in Table~\ref{XYZ}. 

In the following, we compare other thermodynamic properties for the two models. We also compare the results for the Madrid model with those for pure TIP4P/2005 water, but at density values chosen to follow the same temperature-pressure path as shown in Fig.~\ref{modelspvst}. This is conveniently done using the two-state equation of state (EoS) for TIP4P/2005 presented by Biddle~\textit{et al.} in Ref.~\onlinecite{biddle_two-structure_2017}, given that this equation of state is valid at negative pressure and up to $320\,\mathrm{K}$.

\begin{figure}[bbb]
\begin{center}
	\includegraphics[clip,width=0.95\columnwidth]{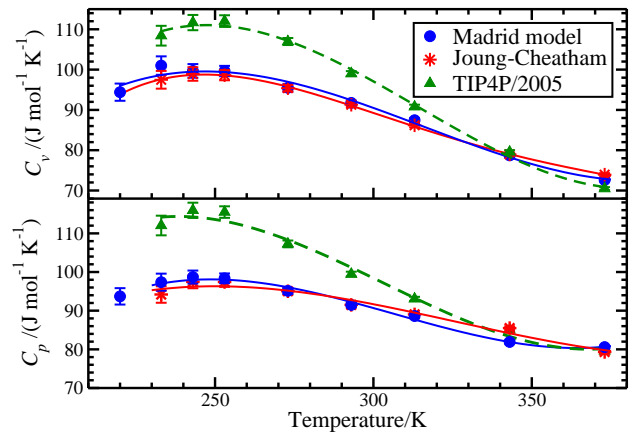}
\end{center}
	\caption{Heat capacity per mole of solution at constant volume (top) and at constant pressure (bottom) along the isochore at $\rho=996.5\,\mathrm{kg\,m^{-3}}$ for Joung-Cheatham (red stars) and Madrid model (blue circles) at $1.323\,\mathrm{mol\,kg^{-1}}$. The green triangles show the values from the EoS for pure TIP4P/2005 water~\cite{biddle_two-structure_2017} along the same temperature-pressure path as followed by the isochore with the Madrid model (see Fig.~\ref{modelspvst}).}
	\label{cvcpbothmodels}
\end{figure}

Figure~\ref{cvcpbothmodels} displays heat capacity at constant volume (top panel) and constant pressure (bottom panel) for the two models. The results are very close to each other. Both models show a mild maximum in $C_\mathrm{V}$ and $C_\mathrm{P}$ around $246-247\,\mathrm{K}$, with a peak being slightly higher for the Madrid model (see also Table~\ref{XYZ}). When comparing the results for pure TIP4P/2005 and for the Madrid model along the same $T-P$ path (Fig.~\ref{modelspvst}), adding salt to water reduces $C_P$ and $C_V$ and makes their peaks less pronounced.
	
\begin{figure}[tt]
	\begin{center}
	\includegraphics[clip,width=0.95\columnwidth]{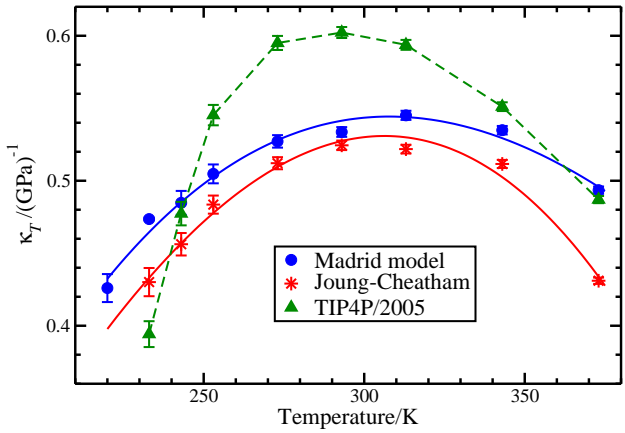}
\end{center}
	\caption{Isothermal compressibility along the isochore at $\rho=996.5\,\mathrm{kg\,m^{-3}}$ for Joung-Cheatham (red stars) and Madrid model (blue circles) at $1.323\,\mathrm{mol\,kg^{-1}}$. The green triangles show the values from the EoS for pure TIP4P/2005 water~\cite{biddle_two-structure_2017} along the same temperature-pressure path as followed by the isochore with the Madrid model (see Fig.~\ref{modelspvst}).}
	\label{kappa-bothmodels}
\end{figure}

Figure~\ref{kappa-bothmodels} displays the isothermal compressibility for the two models as compared to water. The results are again close to each other. Both models show a weak maximum in $\kappa_\mathrm{T}$ around $307-308\,\mathrm{K}$, with the peak being slightly higher for the Madrid model (see also Table~\ref{XYZ}). When comparing the results for pure TIP4P/2005 and for the Madrid model along the same $T-P$ path (Fig.~\ref{modelspvst}), adding salt to water decreases $\kappa_T$ making the maximum less pronounced.
	
	\begin{table}[ttt]
		\caption{Extrema values of the properties along the isochore at $\rho=996.5\,\mathrm{kg\,m^{-3}}$ for Joung-Cheatham and Madrid models at $1.323\,\mathrm{mol\,kg^{-1}}$. 
		These properties (first column) present a maximum/minimum at the temperature and pressure shown in the second and third columns respectively. In the last column, the value of each property at that condition is shown. Heat capacities are given per mole of solution.}
		\label{XYZ}
	\begin{tabular}{cccc}
\hline  
\hline  
        & Joung-Cheatham - TIP4P/2005 &  &\\
\hline
Quantity					&$ T/\mathrm{K}$			& $P/\mathrm{MPa}$	& Value\\
min. $P$			& $260$		& -1515			& \\
		max. $C_{V}$			& $246$ 		& -149.2			& $98.8\,\mathrm{J\,K^{-1}\,mol^{-1}}$\\
		max. $C_{P}$			& $250$		& -149.6			& $96.3\,\mathrm{J\,K^{-1}\,mol^{-1}}$\\
		max. $\kappa_{T}$		& $307$		& -127.1			& $5.3\,10^{-4}\,\mathrm{MPa^{-1}}$\\
		min. $c$			& $293$		& -124.1			& $1390\,\mathrm{m\,s^{-1}}$\\
\hline
\hline
        & Madrid - TIP4P/2005 &  &\\
\hline
Quantity					&$ T/\mathrm{K}$			& $P/\mathrm{MPa}$	& Value\\
min. $P$			& $263$		& -122.3			& \\
		max. $C_{V}$			& $247$		& -118.3			& $99.5\,\mathrm{J\,K^{-1}\,mol^{-1}}$\\
		max. $C_{P}$			& $248$		& -119.1			& $98.1\,\mathrm{J\,K^{-1}\,mol^{-1}}$\\
		max. $\kappa_{T}$		& $308$ 		& -100.4	        & $5.4\,10^{-4}\,\mathrm{MPa^{-1}}$\\
		min. $c$			& $296$		& -98.8			& $1357\,\mathrm{m\,s^{-1}}$\\
\hline
\hline
\end{tabular}
\end{table}

\begin{figure}[tt]
\begin{center}
\includegraphics[clip,width=0.95\columnwidth]{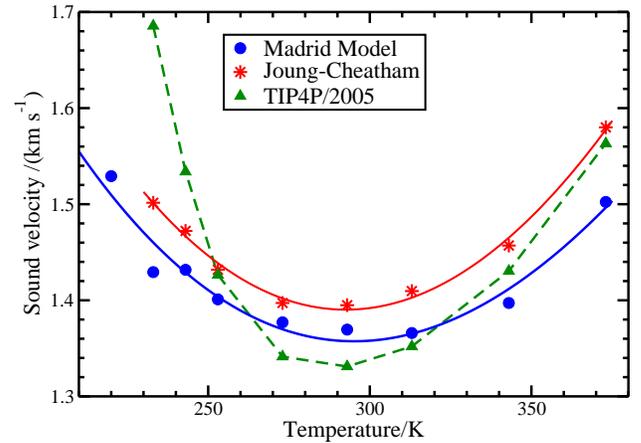}
\end{center}
	\caption{Sound velocity computed along the isochore at $\rho=996.5\,\mathrm{kg\,m^{-3}}$ for Joung-Cheatham (red stars) and Madrid Model (blue circles) at $1.323\,\mathrm{mol\,kg^{-1}}$. The green triangles show the values from the EoS for pure TIP4P/2005 water~\cite{biddle_two-structure_2017} along the same temperature-pressure path as followed by the isochore with the Madrid model (see Fig.~\ref{modelspvst}). Error bars are smaller than the symbol size.}
	\label{sos-bothmodels}
\end{figure}

Finally Fig.~\ref{sos-bothmodels} displays the calculated sound velocity. The two models for salty water give similar results, both with a minimum in $c$ around $293-296\,\mathrm{K}$ (see Table~\ref{XYZ}). The sound velocity is shifted to higher values in the case of the Joung-Cheatham model. When comparing the results for pure TIP4P/2005 and for the Madrid model along the same $T-P$ path (Fig.~\ref{modelspvst}), adding salt to water makes the sound velocity minimum shallower.

From this analysis we conclude that the two NaCl models yield qualitatively identical and quantitatively close results. The most noticeable difference is the magnitude of the pressure reached along the studied isochore. For the remainder of the article, we will present only results obtained with the Madrid model.

\begin{figure}[bbb]
\begin{center}
\includegraphics[clip,height=2.5cm]{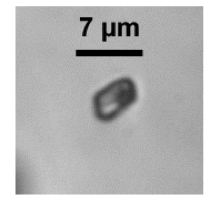}
\end{center}
\caption{Natural sample of fluid inclusion in quartz containing a solution equivalent to $1.20\,\mathrm{mol\,kg^{-1}}$ NaCl molality.}
\label{sample}
\end{figure}

\section{Experimental results\label{sec:exp}}

\subsection{Experimental details}\label{sec:exp_details}

The sample is a natural quartz fragment from the Mont Blanc massif in the French Alps. We have selected a FI containing salt and able to reach a large negative pressure. This particular FI has already been studied, but only above $100\,^{\circ}\mathrm{C}$, in Ref. \onlinecite{mekki-azouzi_brillouin_2015} (sample FI4), which presented a method for using Brillouin spectroscopy as a paleothermometer. Here we have extended the measurements to lower temperatures and more negative pressure, to study the effect of salt on the water anomalies. The sample was cut perpendicular to the c-axis and polished on both sides, resulting in a $200\,\mathrm{\mu m}$ thick slab. The experimental setup was the same as in Refs.~\onlinecite{pallares_anomalies_2014,mekki-azouzi_brillouin_2015,holten_compressibility_2017}. In brief, the sample temperature is controlled within $0.1\,^{\circ}\mathrm{C}$ with a Linkam THMS600 microscope stage. Starting from a state where a bubble is present (biphasic FI), heating the fluid inclusion makes the bubble shrink until the bubble disappears at the homogenization temperature $T_\mathrm{h}$. Cooling down the monophasic fluid inclusion brings it to negative pressure until the sample cavitates and a new bubble appears.
Around $100\,\mathrm{mW}$ of a monomode $532\,\mathrm{nm}$ laser (Verdi Coherent V6) are focused to a $1\,\mathrm{\mu m}$ spot in the inclusion studied, using a Mitutoyo Plan Apo x100 long-working distance objective. The backscattered light is collected through the same objective and routed to a tandem Fabry-Pérot Brillouin spectrometer (JRS Scientific, TFP-1) with entrance and exit pinholes 300 and $450\,\mathrm{\mu m}$, respectively. The spectra are recorded to reach around 300 counts at the Brillouin peak.

This natural sample contains unknown solutes. Sodium chlorine being the most abundant salt found in inclusions, we take as a proxy for the natural solution a NaCl solution. We have estimated its molality from Raman ($1.25\pm 0.05\,\mathrm{mol\,kg^{-1}}$) and from Brillouin ($1.20\pm 0.03\,\mathrm{mol\,kg^{-1}}$)~\cite{mekki-azouzi_brillouin_2015}. Using the latter value, and the measured homogenization temperature $T_\mathrm{h}$, we deduce the density at $T_\mathrm{h}$, $\rho_0 =987.3\,\mathrm{kg\,m^{-3}}$. We account for changes in density with temperature (see Appendix~\ref{Brillouin} for details). Then, using the corresponding refractive index and a viscoelastic analysis of the Brillouin spectra (see Appendix~\ref{Brillouin} for details), we obtain the sound velocity.

\subsection{Comparison between pure and salty water\label{sec:exp-comp}}

We now present the results obtained with the natural salty sample and compare them with those previously obtained with a synthetic pure water sample~\cite{holten_compressibility_2017}.

\begin{figure}[bbb]
\begin{center}
        \includegraphics[clip,width=0.95\columnwidth]{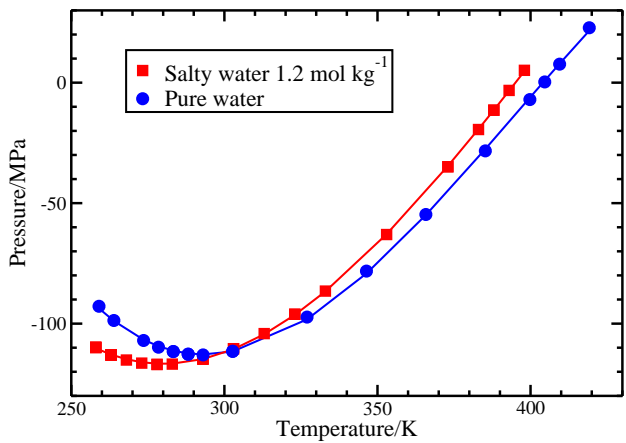}
\end{center}
\caption{Pressure as a function of temperature in the pure (blue circles) and  salty ($1.2\,\mathrm{mol\,kg^{-1}}$ NaCl molality, red squares) samples. The pressure for the pure water sample is obtained from the experimental EoS at negative pressure (Holten~\textit{et al.}~\cite{holten_compressibility_2017}). The pressure for the salty water sample is calculated from the extrapolation of an EoS measured at positive pressure (AlGhafri \textit{et al.}~\cite{al_ghafri_densities_2012}).}\label{fig:Pexp}
\end{figure}

Figure~\ref{fig:Pexp} displays the pressure reached along the experimental path. Rather than a perfect isochore, because of thermal expansion and elasticity of the quartz matrix, the experiment follows a quasi-isochore, with a typical density variation of a fraction of a percent. In the case of the salty sample, the pressure along the quasi-isochore can only be estimated, as it involves the extrapolation of an equation of state for NaCl solutions at positive pressure (AlGhafri \textit{et al.}~\cite{al_ghafri_densities_2012}, see Appendix~\ref{Brillouin} for details). The pressure values for the salty sample should therefore be taken with caution. In contrast, in the case of the pure water sample, the pressure is calculated using the experimental equation of state at negative pressure obtained in Ref.~\onlinecite{holten_compressibility_2017} by thermodynamic integration of a set of speed of sound data at various temperatures and densities. The pressure is therefore more reliable in the pure water case. For the present study, we have taken the data from one of the samples studied in Ref.~\onlinecite{holten_compressibility_2017}, which we selected because its pressure-temperature path was the closest to the one estimated for the present salty sample. This allows a more direct comparison between pure and salty water.

\begin{figure}[bbb]
\begin{center}
    \includegraphics[clip,width=0.95\columnwidth]{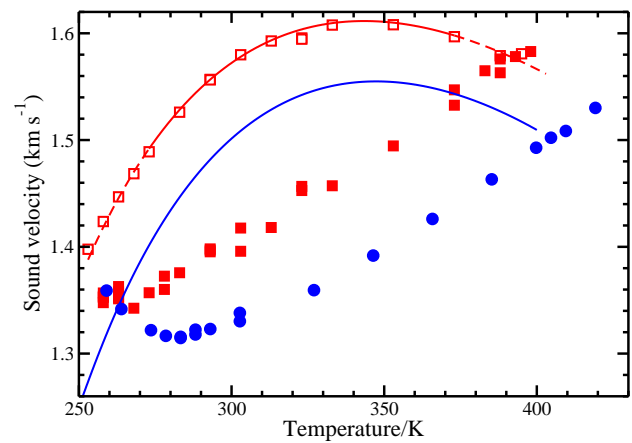}
\end{center}
\caption{Sound velocity as a function of temperature for pure (blue circles) and  salty ($1.2\,\mathrm{mol\,kg^{-1}}$ NaCl molality, red squares) samples. Empty symbols stand for data at the liquid-vapor equilibrium, whereas filled symbols stand for data along the quasi-isochore. Error bars (not shown for clarity) are $6\,\mathrm{m\,s^{-1}}$ (one standard deviation). The blue curve shows the expected sound velocity along the liquid-vapor equilibrium, calculated from the IAPWS EoS~\cite{wagner_iapws_2002}. The solid red curve shows the sound velocity measured at ambient pressure for a $1.2\,\mathrm{mol\,kg^{-1}}$ NaCl molality solution in water~\cite{millero_pvt_1987}; the dashed red lines are extrapolations of these measurements beyond the temperature range in which they were taken.}\label{fig:wexp}
\end{figure}

Figure~\ref{fig:wexp} displays the sound velocity obtained from Brillouin measurements on the pure and the salty samples. When a bubble is present in the fluid inclusion, it is at the liquid-vapor equilibrium. The sound velocity measured under this condition (shown with empty symbols) is in excellent agreement with the values expected from standard measurements for pure and salty water. In the case of the salty samples, Brillouin measurements even provide sound velocity data beyond the previously available limits: they agree with the extrapolation of the literature data (dotted red curves). The sound velocity at liquid-vapor equilibrium increases with salt concentration.

Turning to the comparison between the quasi-isochores for the pure and salty samples, at high temperature they also run parallel to each other, with a $75\,\mathrm{m\,s^{-1}}$ shift. Below $300\,\mathrm{K}$, the pure water data level out before showing a clear minimum at $283\,\mathrm{K}$. The salty water data keeps decreasing when temperature decreases until $268\,\mathrm{K}$, below which one notices a slight increase at the two lowest temperatures. Unfortunately, this increase is comparable to the data scatter, which does not allow us to reach a clear conclusion about the existence of a sound velocity minimum in the case of the salty sample. If the apparent minimum for the salty sample is real, it is located around $15\,\mathrm{K}$ below that for the pure water data.

It turns out that the sound velocity for the two samples becomes equal at the lowest temperature. This means however a stark difference in their behavior: upon cooling, the sound velocity of the pure water sample eventually exceeds the value at liquid-vapor equilibrium, indicating a non-monotonic density dependence of the sound velocity (see Ref.~\onlinecite{pallares_anomalies_2014}, in particular Fig.~4, and Ref.~\onlinecite{holten_compressibility_2017}, in particular Fig.~1). In contrast, the sound velocity for the salty sample always remain below the liquid-vapor equilibrium values in the temperature range studied.

\section{Discussion}
\label{sec:discuss}

\subsection{Comparison between experiments and simulations: quasi-isochores vs. isochores} \label{sec:compar_model-exp}

Experiments on fluid inclusions are performed following a quasi-isochore, whereas simulations are typically performed along a true isochore (see for instance Fig.~\ref{sos-bothmodels}). This may affect the location and even the existence of extrema in thermodynamic quantities. In the case of experiments with pure water, this was carefully taken into account using quartz properties and an iterative procedure~\cite{pallares_anomalies_2014,holten_compressibility_2017}. As samples were studied along different quasi-isochores, interpolation and thermodynamic integration allowed to construct an experimental equation of state. Using this EoS, it was shown that the minimum in sound velocity vs. temperature observed along quasi-isochores remains present along true isochores.

\begin{figure}[bbb]
\begin{center}
        \includegraphics[clip,width=0.95\columnwidth]{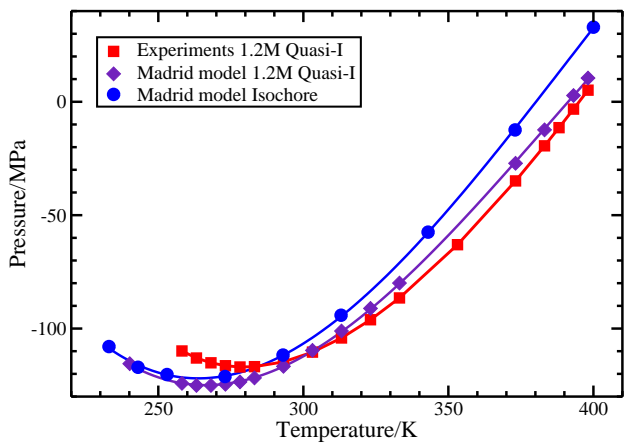}
\end{center}
\caption{Pressure as a function of temperature along the isochore at $\rho=996.5\,\mathrm{kg\,m^{-3}}$ (blue circles) and the quasi-isochore (purple diamonds) for Madrid Model at $1.323\,\mathrm{mol\,kg^{-1}}$. The pressure along the experimental quasi-isochore (red squares, same as Fig.~\ref{fig:Pexp}) is also shown for comparison.}\label{fig:Psimexp}
\end{figure}

\begin{figure}[ttt]
\begin{center}
\includegraphics[clip,width=0.95\columnwidth]{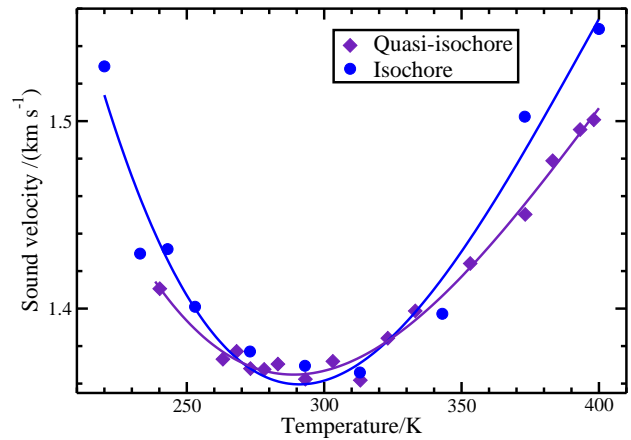}
\end{center}
\caption{Sound velocity along the isochore at $\rho=996.5\,\mathrm{kg\,m^{-3}}$ (blue circles) and the quasi-isochore (purple diamonds) for Madrid Model at $1.323\,\mathrm{mol\,kg^{-1}}$.}
	\label{quasi-vs-iso}
\end{figure}

In the case of salty water, our experiments suggest the existence of a $c$ minimum along the quasi-isochore (Fig.~\ref{fig:wexp}), although it is to shallow to be firmly established at present (see Section III.B). The question arises if the difficulty to conclude about the existence of a minimum might be due to the thermodynamic path. One possibility could be that a stronger minimum exists along an isochore, and becomes smeared out along the quasi-isochore which is actually measured. Unfortunately, data measured along only one quasi-isochore are not sufficient to construct an experimental EoS with enough accuracy and to give the sound velocity along a true isochore, as was done for pure water. Still, the above possibility can be tested with simulations. We repeated the simulations with TIP4P/2005 for water and Madrid model for NaCl at $1.323\,\mathrm{mol\,kg^{-1}}$, along the $\rho=996.5\,\mathrm{kg\,m^{-3}}$ isochore and a quasi-isochore, corresponding to the experimental values for the densities. The experimental densities were calculated taking into account thermal expansion and elasticity of quartz as explained in Appendix~\ref{Brillouin}; this results in a maximum density change from $995.6\,\mathrm{kg\,m^{-3}}$ at $236\,\mathrm{K}$ to $987.1\,\mathrm{kg\,m^{-3}}$ at $399\,\mathrm{K}$. Fig.~\ref{fig:Psimexp} shows that the pressure along the experimental and simulated quasi-isochores are in good agreement. Fig.~\ref{quasi-vs-iso} shows the comparison between sound velocity simulated along the isochore and the quasi-isochore. The changes are small but systematic. They make the sound velocity minimum slightly less pronounced along the quasi-isochoric path than along the isochoric path, but near the minimum the differences are very small.

\begin{figure}[bbb]
\begin{center}
\includegraphics[clip,width=0.95\columnwidth]{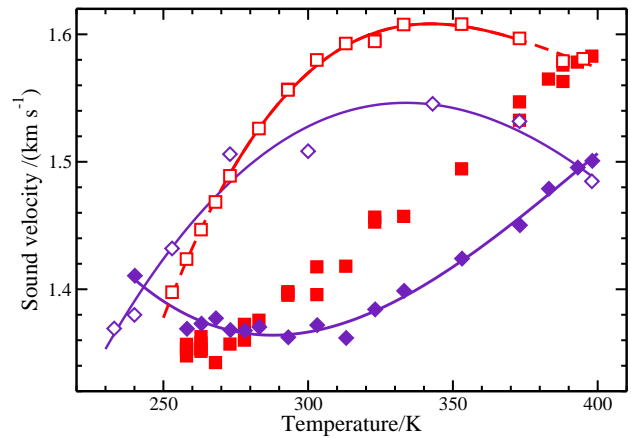}
\end{center}
\caption{Sound velocity along the binodal (empty symbols) and the quasi-isochore (filled symbols) for Madrid model at $1.323\,\mathrm{mol\,kg^{-1}}$ (purple diamonds) and experiments ($1.2\,\mathrm{mol\,kg^{-1}}$ NaCl molality, red squares).}
\label{sos}
\end{figure}

In the case of experiment, we conclude that it is unlikely that measuring along a true isochore (if this were possible) rather than along a quasi-isochore would qualitatively change our results. Fig.~\ref{sos} shows the comparison of the sound velocity between experiments and simulations. The rather close agreement along the quasi-isochore suggests that a minimum could exist in experiments, although, as noted in Section~\ref{sec:exp}, the scatter of experimental data does not allow to reach a clear conclusion. To decide about the existence or absence of a minimum in experiments, a critical step forward would be to acquire more data at lower temperature. In fact, the sample remains in the fluid state at lower temperature, as confirmed with Raman spectroscopy for instance. Unfortunately, we were not able take sound velocity data below $258.15\,\mathrm{K}$ because sound attenuation then becomes too strong and we loose the Brillouin signal.

\subsection{Origin of the minima in sound velocity for pure water: LLCP vs. spinodal\label{sec:Altabet}}

A major question in the context of the debate about the origin of water's anomalies is the origin of the minima in sound velocity. In the case of pure water, such minima have been observed both in experiments and simulations~\cite{pallares_anomalies_2014,holten_compressibility_2017}. In Ref.~\onlinecite{holten_compressibility_2017}, several first-principle, thermodynamic relations were derived, and their consequences on the different lines of anomalies were studied. In particular, it was shown that, if a LLCP exists, a line of sound velocity minima along isobars (Lm$c|P$) must emanate from this critical point. This makes the existence of a Lm$c|P$ a necessary (but not sufficient) condition for the second critical point scenario~\cite{poole_phase_1992} to be valid. If a LLCP exists in pure water, it will be preserved by the addition of a solute. However, the critical behaviour differs~\cite{sengers_critical_1994}: the osmotic susceptibility, not the isothermal compressibility, diverges at the LLCP, from which a line of osmotic susceptibility maxima emerges. Still, for sufficiently low concentrations, a line of isothermal compressibility maxima will be preserved; the smaller the solute concentration, the closer the lines of isothermal compressibility and osmotic susceptibility maxima. We thus expect that the existence of a LLCP in pure water would still cause a line of sound velocity minima in salty water, with minima that would become less and less pronounced with the increase of salt concentration. This is consistent with what is observed in our simulations (see Section~\ref{sec:compar_models}).

However, Altabet~\textit{et al.}~\cite{altabet_thermodynamic_2017} recently investigated another possible source of sound velocity minima along isochores: the case of a liquid-vapor spinodal exhibiting a maximum in its density vs. temperature. These authors showed that ``a maximum spinodal density in water results in a locus of maximum compressibility and a minimum speed of sound that are independent from any influence of a LLCP ''. They reproduced our previous results for the sound velocity minima along two isochores obtained with TIP4P/2005~\cite{pallares_anomalies_2014}, and obtained more minima for isochores at lower densities. They argued that ``the $\kappa_T^\mathrm{max}$ [line of $\kappa_T$ maxima along isochores] is not the negative pressure extension of a line emanating from higher pressure. Instead, it is due to the peculiar behavior of water’s spinodal in its $T-\rho$ phase diagram and originates at
negative pressure.''

\begin{figure}[ttt]
	\includegraphics[clip,width=0.95\columnwidth]{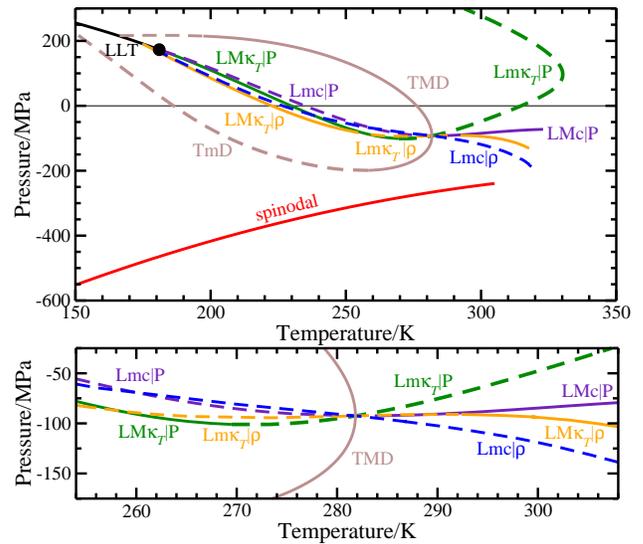}
	\caption{Remarkable lines in the phase diagram of TIP4P/2005. The black curve shows the LLT, terminated at the LLCP (black dot), and the red curve the spinodal. Lines of maxima and minima along isobars ($|P$) are shown with solid and dashed curves, respectively, for density (brown), $\kappa_T$ (green), and $c$ (purple). Lines of maxima and minima along isochores ($|\rho$) are shown with solid and dashed curves, respectively, for $\kappa_T$ (orange), and $c$ (blue). The bottom panel shows a close-up around the TMD turning point.}
\label{lines2005}
\end{figure}

A question thus arises about the relevance of our observations to the debate about the existence of a LLCP. In the case of pure water, the answer to this subtle question can be found by looking at the lines of extrema for TIP4P/2005, which can be plotted to their full extent in Fig.~\ref{lines2005} thanks to the available parameterization of the simulation data with a two-state model~\cite{biddle_two-structure_2017}. As explained in details in the Supporting Information of Ref.~\onlinecite{holten_compressibility_2017}, there are thermodynamic requirements. For a quantity $X$, let us call Lm$X|Y$ and LM$X|Y$ the loci of $X$ minima and maxima along a constant $Y$ path. A Lm$c|P$ must emanate from the LLCP. Let us call M the point at which the temperature of maximum density (TMD) reaches its maximum. At M, an extremum in sound velocity along isobars must be reached. If this extremum is a minimum, the Lm$c|P$ extends from the LLCP to temperatures above the TMD maximum. If, instead, the extremum at M is a sound velocity maximum, then the Lm$c|P$ will stop at a temperature lower than that of M, and will merge with a LM$c|P$. TIP4P/2005 is a borderline case, with the Lm$c|P$ connecting with the LM$c|P$ nearly exactly at M. Coming back to the argument by Altabet~\textit{et al.}, we see in Fig.~\ref{lines2005} that they plotted in their Fig.~4 only the high temperature part of the LM$\kappa_T|\rho$. Upon cooling, it becomes a line of minima in $\kappa_T$ along isochores, Lm$\kappa_T|\rho$, and upon further cooling, once again a LM$\kappa_T|\rho$. The low temperature LM$\kappa_T|\rho$ eventually tracks the low temperature part of the Lm$c|\rho$, the two lines passing to the left of the LLCP, as they should~\cite{holten_compressibility_2017}. At low temperature, LM$\kappa_T|\rho$ and Lm$c|\rho$ become nearly parallel to LM$\kappa_T|P$ and Lm$c|P$, respectively, the two latter lines terminating at the LLCP. We see that, if at high temperature LM$\kappa_T|\rho$ and Lm$c|\rho$ are indeed caused by the spinodal, at low temperature they are caused by the LLCP.

The key to recognize the possible cause for these anomalies is to scrutinize their location relative to the TMD. To the right of (i.e. at temperatures above) the TMD they are influenced by the spinodal, whereas to the left of (i.e. at temperatures below) the TMD they are related to the LLCP. Altabet~\textit{et al.} are right about the spinodal origin of the minimum sound velocity along the $933.2\,\mathrm{kg\,m^{-3}}$ isochore for TIP4P/2005 water, as it lies to the right of the corresponding TMD. In our previous work\cite{pallares_anomalies_2014}, we simulated this particular isochore to match the experimental density, but we were not aware of the issue at that time. However, the experimental equation of state deduced from our measurements puts the observed minima in sound velocity along isochores (for 6 samples at different densities, including $933.2\,\mathrm{kg\,m^{-3}}$) to the left of the TMD. This also explains why corresponding sound velocity minima along isobars are observed in the experimental EoS (see Fig.~2 of Ref.~\onlinecite{holten_compressibility_2017}), which would not be the case if they were located to the right of the TMD. The experimental findings for pure water thus realize the necessary (but not sufficient) condition for a LLCP to exist in real water.

\subsection{Other force-fields}

\begin{figure}[bbb]
	\includegraphics[clip,width=0.95\columnwidth]{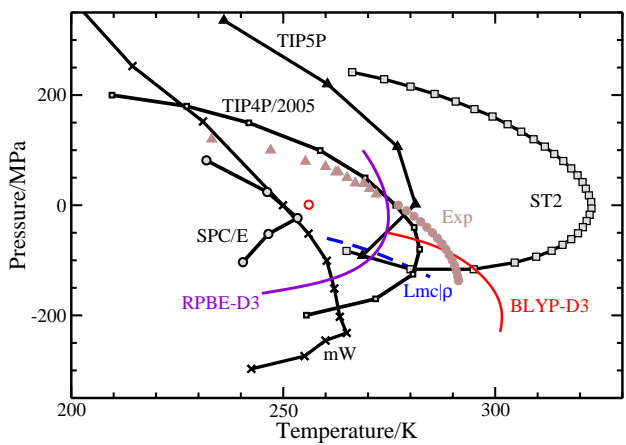}
\caption{TMD lines computed for different models presented in Ref.~\onlinecite{singraber_density_2018}. Brown symbols represent the experimental measurements for pure water at negative (circles \cite{holten_compressibility_2017}) and positive (triangles \cite{holten_equation_2014}) pressures. The blue dashed curve shows the location of experimental $c$ minima along isochores~\cite{holten_compressibility_2017}.} 
\label{fig:TMD}
\end{figure}

Now that, thanks to Altabet~\textit{et al.}, we have realized the importance of the TMD location, we can wonder if we would have been able to provide a better comparison between experiments and simulations. For TIP4P/2005, we selected the same density values for the simulation as in the experiment, which yielded a good agreement for the sound velocity~\cite{pallares_anomalies_2014}, but the resulting $c$ minima along isochores lied to the right of the TMD. Simulating TIP4P/2005 isochores at a sufficiently higher density would place the sound velocity minimum to the left of the TMD, but the agreement with experimental values for sound velocity and pressure would decrease. This issue arises from the fact that, for TIP4P/2005, the TMD changes slope around $-80\,\mathrm{MPa}$ and $\rho \simeq 955 \,\mathrm{kg\,m^{-3}}$, whereas in experiments\cite{holten_compressibility_2017}, the TMD keeps a negative slope to more negative pressures (at least $-137\,\mathrm{MPa}$), and a sound velocity minimum is observed only for $\rho < 951 \,\mathrm{kg\,m^{-3}}$. Therefore, we may consider using another force-field to make the comparison more meaningful. In a recent work, Singraber and Dellago have compiled the available TMD lines and added two new ones based on ab initio trained high-dimensional neural network potentials\cite{singraber_density_2018}. We compare them to the experimental location of the Lm$c|\rho$ and  Lm$c|P$ in Fig.~\ref{fig:TMD}. The experimental lines lie in an appropriate position relative to the simulated TMDs only for TIP4P/2005 and BLYP-D3. However, the TMD for BLYP-D3 is not satisfactory as it does not reach positive pressure. Therefore TIP4P/2005 is the best possible choice among the potentials compared in Fig.~\ref{fig:TMD}. 

\subsection{Origin of the minima in sound velocity for salty water, and perspective for future experiments}

\begin{figure}[bbb]
\includegraphics[clip,width=0.95\columnwidth]{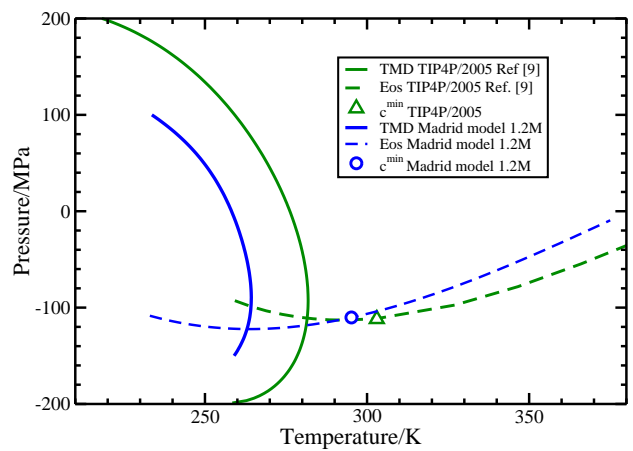}
	\caption{
	TMD computed for pure water~\cite{biddle_two-structure_2017} (solid green curve) and Madrid model at $1.323\,\mathrm{mol\,kg^{-1}}$ (solid blue curve). The sound velocity minima are shown with a green triangle and a blue circle, respectively, along the corresponding isochores (green and blue dashed curve, respectively).}
\label{fig:TMDsalty}
\end{figure}

Let us now come back to the case of salty water. For simulations with TIP4P/2005 for water and the Madrid model for salt, we chose a density which gives a pressure close to the experimental one (Fig.~\ref{fig:Psimexp}). The sound velocity along this isochore shows a clear minimum (Fig.~\ref{sos-bothmodels}). In the light of the work by Altabet~\textit{et al.}, we now need to discuss the location of the $c$ minimum with respect to the TMD. We have computed the TMD from several NVT simulations at the same NaCl concentration. The result is displayed in Fig.~\ref{fig:TMDsalty}. As for pure TIP4P/2005, the minimum in sound velocity along the simulated isochore lies to the right of the TMD, and is therefore due to the spinodal.

What can we say about the experimental case? First, the existence of the $c$ minimum needs to be confirmed (see Section~\ref{sec:exp-comp}). Still, assuming it exists, would it be due to the spinodal? As shown for pure water in Section~\ref{sec:Altabet}, the $c$ minima in simulations and experiments along the same isochores might not have the same origin, because they can be located on opposite sides of the corresponding TMD. Therefore, the knowledge of the experimental TMD for the salt concentration studied is required. Unfortunately, it cannot be obtained from the measurement in a single FI. To give a crude estimate, we resort to an extrapolation of an EoS for NaCl solutions measured at positive pressure\cite{al_ghafri_densities_2012}: at $-115\,\mathrm{MPa}$, the pressure estimated at the condition of the shallow $c$ minimum, the extrapolation gives a TMD at $279\,\mathrm{K}$. The $c$ minimum would thus not be due to the spinodal.

However, these extrapolations are highly uncertain. This highlights the need for more measurements. Measuring the same salty sample at lower temperatures would help in deciding about the existence of a $c$ minimum; however, this is challenging because of the strong sound attenuation which makes the Brillouin peak disappear. It would be interesting to measure samples with a lower salt concentration, for which we expect a behaviour between the present salty sample and pure water, and therefore the possibility of observing a sharper minimum. For a given concentration, a more systematic study with measurements of samples at various densities, as was done for pure water\cite{holten_compressibility_2017}, is also desirable. This will allow determining the experimental equation of state, and, if a $c$ minimum is confirmed, elucidate its location with respect to the TMD.

\section{Conclusion}
We have measured and simulated sound velocity in a salty aqueous solution at negative pressure. For the chosen conditions, the simulations display a minimum in sound velocity vs. temperature along an isochore. In the experimental case, data suggests the existence of a minimum, although it cannot be ascertain at present in view of the data scatter. The respective location of sound velocity minima and line of density maxima in the studied solutions tells us if the minima are due to the spinodal or not. Further experimental work is needed, using for each salt concentration several samples at various densities, to determine the corresponding line of density maxima at negative pressure and look for sound velocity minima. There is also room for improving force fields for pure water, in order to correctly reproduce the experimental line of density maxima from positive to negative pressure.

\acknowledgments We thank Mikhail A. Anisimov for helpful discussions on critical phenomena in solutions, and Christoph Dellago and Andreas Singraber for providing the data used to plot Fig.~\ref{fig:TMD}. C.S.P.T. and F.C. acknowledge funding by the
European Research Council under the European Community’s FP7 Grant Agreement 240113. CV acknowledges fundings from the  Spanish Ministry of Education FIS2016-78847.
AZ was funded by CONACYT (PhD fellowship) and MAG by the Spanish Ministry of Education (Juan de la Cierva fellowship).

\appendix

\section{Obtaining sound velocity from the Brillouin spectra}\label{Brillouin}
The refractive index is needed to obtain the sound velocity from the Brillouin shift. We calculate the refractive index $n$ as follows. We assume the validity of the Gladstone-Dale relation, $n(T,m) = 1 + K(m)\rho(T,m)$ where $\rho(T,m)$ is the density at temperature $T$ and molality $m$. We calculate the constant $K(m=1.20)$ using interpolated values at 20$\,^{\circ}$C for the density and refractive index (at $589\,\mathrm{nm}$)~\cite{noauthor_concentrative_2014}. To compute $n(T,m=1.20)$ at various temperatures, we use the correlation of density measurements for NaCl solutions in the range 25 to 200~$^{\circ}$C, 0 to $6\,\mathrm{mol^,kg^{-1}}$, and $0.1$ to $68.5\,\mathrm{MPa}$~\cite{al_ghafri_densities_2012}. For simplicity we took $P=0.1\,\mathrm{MPa}$ when a bubble was present in the inclusion. Note that we used the refractive index tabulated at $589\,\mathrm{nm}$, whereas the experiment is carried out at $532\,\mathrm{nm}$. In our experiments, this difference was not significant as we checked by measuring Brillouin spectra as a function of temperature and molality for NaCl solutions contained in capillaries (Fig. 4b of Ref.~\onlinecite{mekki-azouzi_brillouin_2015}). As the exact solute present in the natural sample is unknown, we did not attempt to repeat the procedure at the correct wavelength. In future work, when synthetic samples containing only NaCl and water will be used, a more accurate analysis would be in order.

The Brillouin spectra are analyzed with the viscoelastic model, convoluted with the instrumental response function (see Ref.~\onlinecite{pallares_anomalies_2014} for details). All fits are excellent with a typical reduced $\chi^{2}$ around 1 (at most 1.7). We use a constant sound velocity at infinite frequency $c_{\infty}=3000\,\mathrm{m\,s^{-1}}$, as was done for pure water. The analysis yields the sound velocity at zero frequency $c_{0}$. All results for $c_{0}$ are then multiplied by a common correction factor 1.01029. The correction factor was determined from the ratio between the expected value for pure water\cite{wagner_iapws_2002} and the raw $c_{0}$ obtained for pure water in a capillary. This includes possible biases in the scan amplitude and in the definition of the collection angle. No significant variation of the correction factor was observed between 20 and 60 $^{\circ}$C, therefore the average ratio was used as a constant correction factor. In the regime of the experiment, ($c_{0}/c_{\infty})^{4} (2\pi \Delta f_\mathrm{B} \tau)^{2} \ll 1$, where $\Delta f_\mathrm{B}$ is the Brillouin frequency shift and $\tau$ the viscoelastic relaxation time), the correction for $c_{0}$ simply amounts to multiplying all raw results for $c_{0}$ by a constant factor 1.01029.

For the monophasic inclusions, a first analysis is carried out assuming the density remains constant, equal to $\rho_0 =987.3\,\mathrm{kg\,m^{-3}}$ determined from the homegenization temperature $T_\mathrm{h}$. According to the Gladstone-Dale relation, the refractive index is also constant. Then two corrections are needed to account for the change in volume of the inclusion. The first arises from the thermal expansion of quartz, the second from its elasticity. A rough estimate of the pressure is obtained by extrapolating to negative pressure the correlation from AlGhafri \textit{et al.}~\cite{al_ghafri_densities_2012} The correlation was developed for pressures in the range 0.1 to $68.5\,\mathrm{MPa}$, but it is well behaved to large negative pressures. At each temperature, the pressure in the inclusion is estimated as the extrapolated pressure $P$ at which the density would be equal to $\rho_{0}$. A new density $\rho_\mathrm{new}$ is then obtained accounting for quartz expansion and stretching, using:
\begin{equation}
\rho_\mathrm{new} = \rho_{0} \left(1+ \alpha_\mathrm{V} (T-T_\mathrm{h}) + \frac{1+\nu}{1-2\nu}\frac{P}{2B}\right)^{-1} \; ,
\end{equation}
where $\alpha_\mathrm{V}$, $\nu$ and $B$ are the volume expansion coefficient, Poisson coefficient, and bulk modulus of quartz, respectively. Values are given in Ref.~\onlinecite{pallares_equation_2016}.

The Gladstone-Dale relation gives the refractive index corresponding to $\rho_\mathrm{new}$ , which is then used to obtain the value of $c_{0}$ corrected for non-isochoric effects. These corrections gives only minor changes to the sound velocity (at most 2.5 m/s or 0.19 \% at -15 $^{\circ}$C, to compare to our 0.4 \% uncertainty). Note that, in previous work on pure water, the measured sound velocity itself was used to obtain an experimental equation of state at negative pressure, and hence the pressure in the inclusion. The procedure was iterated until convergence was achieved (which takes only 2 to 3 iterations). In view of the present data limited to one sample, and on the very moderate correction calculated, we choose not to use the iterative procedure and to limit the density correction to the first-order approximation presented above. As the possible resulting error is on the density correction, we expect the values of density, refractive index and sound velocity to be rather accurate. In contrast, the pressure itself, displayed in Fig.~\ref{fig:Pexp}, is more sensitive to the choice of extrapolation for the equation of state.

\section*{References}


%

\end{document}